\documentclass[11pt,showpacs,preprintnumbers,amsmath,amssymb,prd,nofootinbib,superscriptaddress]{revtex4-2}

\usepackage[colorlinks=true,pdfstartview=FitV,linkcolor=blue,citecolor=blue,urlcolor=blue,breaklinks=true]{hyperref}
\usepackage{graphicx}
\usepackage[T1]{fontenc} 
\usepackage{amssymb,amsmath,bm,natbib}
\usepackage{color}
\usepackage{slashed}
\usepackage{graphics}
\usepackage{graphicx}
\usepackage[utf8]{inputenc}
\usepackage[caption=false]{subfig}
\usepackage{hyperref}
\usepackage{url}
\usepackage{dsfont}
\usepackage{float}
\usepackage{cancel}
\usepackage{units}
\usepackage{blindtext}
\usepackage[utf8]{inputenc}
\usepackage{upgreek}
\usepackage{booktabs}
\usepackage[dvipsnames,table,xcdraw]{xcolor}
\usepackage{enumerate}
\usepackage{mathtools}
\usepackage{soul,color}
\usepackage[normalem]{ulem}

\newcommand{\orcid}[1]{\href{https://orcid.org/#1}{\includegraphics[width=10pt]{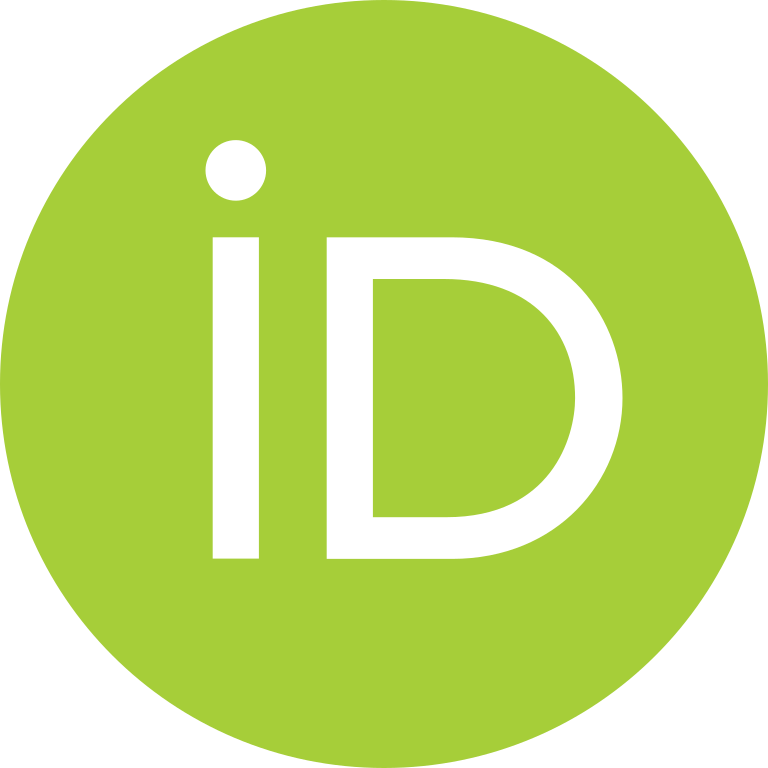}}}

\begin{document}

\title{ Causality Violating Solutions in Curvature-Squared Gravity}

\author{J. C. R. de Souza  \orcid{0000-0002-7684-9540}}
\email{jean.carlos@fisica.ufmt.br }
\affiliation{Programa de P\'{o}s-Gradua\c{c}\~{a}o em F\'{\i}sica, Instituto de F\'{\i}sica,\\ 
Universidade Federal de Mato Grosso, Cuiab\'{a}, Brasil}

\author{A. F. Santos \orcid{0000-0002-2505-5273}}
\email{alesandroferreira@fisica.ufmt.br}
\affiliation{Programa de P\'{o}s-Gradua\c{c}\~{a}o em F\'{\i}sica, Instituto de F\'{\i}sica,\\ 
Universidade Federal de Mato Grosso, Cuiab\'{a}, Brasil}

\author{R. Bufalo \orcid{0000-0003-1879-1560}}
\email{rodrigo.bufalo@ufla.br}
\affiliation{Departamento de F\'{\i}sica, Universidade Federal de Lavras,\\
37203-202, Lavras, Minas Gerais, Brazil}

\begin{abstract}
In this paper, we consider some causality violating solutions in the curvature-squared gravity in order to examine whether closed timelike curves (CTCs) are allowed in these models. 
These aspects are studied in terms of the G\"odel, G\"odel-type and axially symmetric cosmological solutions.
We observe that the G\"odel and G\"odel-type metrics are causal solutions of the model so that CTCs are now allowed and, surprisingly, every contribution involving the Weyl tensor is removed from the solutions.
Hence, in order to study the effect (if any) of the Weyl tensor (an conformal symmetry) into CTCs a third metric is considered.
In this case, we obtain contributions due to the Weyl tensor to the energy density and led to modifications of the weak energy condition. 
\end{abstract}

\maketitle

\section{Introduction}
\label{sec1}

In contrast to the cosmological isotropy of the standard $\Lambda$CDM model \cite{Saadeh:2016sak}, which has been supported by the precision measurements of the Planck satellite, recent observations from the
James Webb Space Telescope (JWST) have reported an
unexpected alignment of the axes of rotation of galaxies throughout the observable universe \cite{Shamir:2025rzb}.
This violation of cosmological isotropy has been ascribed to an inherent rotation of the universe \cite{Shamir:2025rzb}.
Although current understanding supports an isotropic Universe with negligible rotation, investigations continue to challenge and test this foundational assumption.

On the phenomenological and theoretical side, we have seen a renewed attention to rotating cosmologies in recent years.
An analysis of the WMAP data has indicate that a rotation could explain observed temperature fluctuations within a Bianchi VII(h) cosmological framework \cite{McEwen:2005bm}.
Moreover, a rotating cosmology has been suggested as a
possible method of resolving the Hubble tension \cite{Szigeti:2025jxz}.
Another interesting aspect about (global) rotational spacetimes is the study of black hole solutions, which allowed enhance the understanding of causality and thermodynamics, and also provided insights into the compatibility
of horizons and global cosmic rotation, revealing rich structures and dynamical behaviors \cite{Gimon:2003ms,Wu:2007gg,Pantig:2025kxm}.

One of the most known rotational cosmological model was proposed by K. G\"odel, showing that  General Relativity does not fundamentally exclude cosmological solutions exhibiting global rotation \cite{Godel:1949ga}.
Actually, this model is endowed with exotic causal structures, allowing the presence closed timelike curves (CTCs) \footnote{The existence of CTCs in a given spacetime, means that no ideal and unique Cauchy surface can be defined, or more moderately, the spacetime might contain open sets in which geodesics incompleteness occurs in the case of future timelike singularities.}, which challenged prevailing conceptions of causality and temporal ordering.
Moreover, a first generalization of this solution to a wider class is known as G\"odel-type metrics, which fulfils the conditions of spacetime homogeneity \cite{Reboucas:1982hn,Teixeira:1985wd,Reboucas:1986tz}.
A second modification now considered an extension of G\"odel's cosmology which does allow for an expanding as well as rotating universe \cite{Obukhov:2000jf}.

Since G\"odel's solutions and its generalizations have successfully been applied in interesting phenomenological cosmological studies, it is worth to consider them also in modified gravitational theories \cite{Sotiriou:2008rp,Nojiri:2010wj,Nojiri:2017ncd,Petrov:2020wgy}.
Although G\"odel-type metrics properties have been studied in some modified gravity models \cite{Gama:2017eip,Nascimento:2023zok,Silva:2024kix}, none of them have considered the case of Weyl's gravity.
 
Weyl gravity is known to be invariant under local conformal transformation of the metric, $g_{\mu\nu} \to  \Omega^2 (x)g_{\mu\nu}$ \cite{Weyl:1918pdp,Mannheim:2011ds}, being insensitive to angles.
Moreover, among its interesting properties, we can observe that it is a power-counting renormalizable theory of gravity due to the presence of higher-order derivatives in the Lagrangian, being considered as an ultraviolet completion of gravity \cite{Kaku:1982xt,Antoniadis:1984kd,boulware,Kluson:2013hza,tHooft:2010xlr,Maldacena:2011mk,Park:2012ds,Stelle:1976gc,Fradkin:1981iu}.
In recent years, black hole solutions, its thermodynamics and observational tests have been studied in Weyl gravity \cite{Asuncion:2025cxw,Khodadi:2025gtq,Lessa:2025sku,Sarmiento:2025qdl}.

In addition, it is worth to remark that from a mathematical point of view, considering Tipler’s theorems \cite{Tipler:1977eb} and also the definition of CTCs, the presence of CTCs and timelike singularities might have similar implications, that the boundary of the spacetime is a compact vicious \cite{Sanchez,Oikonomou:2018qsc}.
Moreover, given the above relation and our interest in a modified gravity perspective, it is well known that $R^2$ terms can be sufficient to removes finite time singularities \cite{deHaro:2023lbq,Odintsov:2018uaw,Oikonomou:2020oex}.
Hence, based on the above facts, it is very natural to expect that the Weyl gravity have an interesting effect into the existence or not of CTCs.

Therefore, the main proposal of this work is to examine the possibility violation of causality and temporal ordering in the most general parity even and diffeomorfism invariant quadratic curvature gravity, and its relation with the conformal sector (in terms of the Weyl tensor).
 This work is organized as follows: Section \ref{sec2} provides preliminary discussions and revise the main aspects of the model and field equations.
In Section \ref{sec3}, we consider the original G\"odel universe in this quadratic curvature model and discuss the existence of physical solutions and its causal structure.
Moreover, in Section \ref{sec4} we examine the G\"odel-type metrics in order to shed light about the relation of the physical solutions and causal structure with the conformal sector.
Since the previous models' solutions were insensitive to the conformal Weyl effects, in Section \ref{sec5} we consider a third model to establish the contributions of the conformal sector into the causal structure of the solutions.
Finally, our final remarks are presented in Section \ref{sec6}.

\section{ Quadratic curvature terms gravity}
\label{sec2}

The gravitational action that describes parity even and generally covariant quadratic curvature terms is written as
\begin{eqnarray}
\label{1}
S=\int d^4x\,\sqrt{-g}\left[\frac{\Lambda}{\kappa}+\frac{R}{2\kappa}-\frac{\alpha}{4}C_{\mu\nu\rho\sigma}C^{\mu\nu\rho\sigma}+\frac{\beta}{8}R^2+\gamma G\right]
+S_m, 
\end{eqnarray}
where $\kappa$, $\alpha$, $\beta$, and $\gamma$ are coupling constants, $\Lambda$ is the cosmological constant and $R$ is the Ricci scalar. Matter is assumed to be coupled minimally to the metric of spacetime.
The Weyl tensor is by definition the traceless part of the Riemann tensor, which in $d$-dimensions reads
\begin{eqnarray}
\label{3}
 C_{\rho \mu \nu \sigma} =  R_{\rho \mu \nu \sigma} - \frac{2}{(d-2)} \left( g_{\rho [\nu}R_{\sigma] \mu} - g_{\mu [\nu } R_{\sigma] \rho}\right)   + \frac{2}{(d-2)(d-1)}   g_{\rho [ \nu} g_{\sigma] \mu} R.
\end{eqnarray}
Moreover, the $\gamma$ dependent term in the action \eqref{1} is proportional to the Gauss-Bonnet curvature term
\begin{equation}
G=R_{\alpha\beta\gamma\delta}R_{\mu\nu\rho\sigma}\epsilon^{\alpha\beta\mu\nu}\epsilon^{\gamma\delta\rho\sigma}=R_{\mu\nu\rho\sigma}R^{\mu\nu\rho\sigma}-4R_{\mu\nu}R^{\mu\nu}+R^{2}
\end{equation}
whose integral, in four dimensions, becomes a topological invariant. Therefore, under smooth variations of the spacetime geometry that do not alter its topology, the Gauss-Bonnet contribution to the action can be neglected.

For different values of the parameters $(\kappa,\alpha,\beta,\gamma)$ the action \eqref{1} displays intriguing behavior, modifying its degrees of freedom \cite{Kluson:2013hza}, being a renormalizable field theory \cite{Stelle:1976gc}, possess the property of asymptotic freedom \cite{Fradkin:1981iu},  etc.

By varying the action \eqref{1} with respect to the metric, the field equations are given by
\begin{align}
\label{2}
& -g_{\mu \nu}\frac{\Lambda}{\kappa} + \frac{1}{\kappa}\left( R_{\mu \nu}-\frac{1}{2}g_{\mu \nu} R\right) + \alpha \left(2\nabla ^{\rho} \nabla ^{\sigma} C_{\rho \mu \nu \sigma} + C_{\rho \mu \nu \sigma}R^{\rho \sigma} \right)   \cr
 &  + \frac{\beta}{2}\left[ R\left( R_{\mu \nu} - \frac{1}{4} g_{\mu \nu}R \right)- \nabla _{\mu} \nabla _{\nu } R + g_{\mu \nu} \nabla ^{\rho} \nabla _{\rho} R \right] = T_{\mu \nu} ,
\end{align}
in which the energy-momentum tensor of matter is defined as
\begin{equation}
T_{\mu\nu} = -\frac{2}{\sqrt{-g}}\frac{\delta S_m}{\delta g^{\mu\nu}}.
\end{equation}

Although some aspects of acausal structures in Gödel-type solutions have been investigated in certain classes of higher-order gravity \cite{Gama:2017eip, Nascimento:2023zok}, no attention has been given to the case of Weyl gravity.
Hence, our main interest in the present analysis is to determine whether and how the Weyl tensor (and consequently the conformal symmetry) modifies the causal structure of some metric solutions that are known to allow the presence of closed timelike curves (CTCs).

\section{G\"{o}del Universe}
\label{sec3}

The metric that describes the G\"{o}del universe, a rotating, homogeneous and isotropic (but non-expanding) solution of GR, is given by the line element \cite{Godel:1949ga,Reboucas:1982hn}
\begin{eqnarray}
\label{6}
ds^2=a^2\left(\left(dt-\mathcal{H}\left(x\right)dy\right)^{2}-dx^{2}-\mathcal{D}\left(x\right)dy^{2}-dz^{2}\right),
\end{eqnarray}
where $\mathcal{H}\left(x\right) = e^x$ and $\mathcal{D}\left(x\right) = \frac{e^{2x}}{2}$, and $a$ is a positive number related with the angular velocity (vorticity) of the surrounding matter about the $y$-axis.

In order to solve the field equations \eqref{2}, we shall compute some important geometric quantities, including the non-zero Ricci tensor components
\begin{eqnarray}
\label{9}
     R_{00} = 1,\quad\quad R_{02} = e^x, \quad\quad R_{22} = e^{2x}
\end{eqnarray}
while the scalars associated with the G\"{o}del metric are given by
\begin{eqnarray}
\label{11}
R &=& g^{\mu \nu}R_{\mu \nu} = g_{\mu \nu}R^{\mu \nu} = \frac{1}{a^2} ,\\ \label{12}
 R_{\mu \nu}R^{\mu \nu} &=& \frac{1}{a^4}.
  \label{13}
\end{eqnarray}
Thus, since $R$ does not depend on any of the variables $(t,x,y,z)$, but only on the constant $a$ we have that the derivative terms $\nabla_{\mu} \nabla_{\nu} R$ and $g_{\mu \nu} \nabla^{\rho} \nabla_{\rho} R$ do not contribute.

Moreover, considering the Weyl tensor contributions for the G\"{o}del metric, we  define by simplicity of notation
\begin{equation}
\label{4}
   B_{\mu \nu} = \nabla ^{\rho} \nabla ^{\sigma} C_{\rho \mu \nu \sigma} \, ,
\end{equation}
so that its components read
\begin{align}
\label{14}
B_{00}&=e^{-x}B_{02}=\frac{4}{5}e^{-2x}B_{22}=\frac{3\left(d-3\right)}{a^{2}\left(d-2\right)},\\
  \nonumber B_{11}&= \frac{8 d^2-33 d+27}{2 a^2 (d-2) (d-1)}.
\end{align}
The remaining contributions from the Weyl tensor come from the coupling
\begin{equation}
\label{5}
   Z_{\mu \nu} =  C_{\rho \mu \nu \sigma}R^{\rho \sigma} \, , 
\end{equation}
whose components yields
\begin{eqnarray}
\label{15}
    \nonumber 2 Z_{11} = - Z_{33} =  \frac{1}{3 a^2},\quad  Z_{22}= \frac{e^{2 x}}{12 a^2}.
\end{eqnarray}
Since the contributions from the Weyl tensor do not vanish, we can observe that the G\"odel universe are not conformally flat.

For simplicity, we shall gather the geometric contributions (left-hand side) of Eq.\eqref{2} in terms of $\mathbb{J}_{\mu \nu}$, so that its non-vanishing components are given by
\begin{eqnarray}
\label{16}
    \nonumber \mathbb{J}_{00}&=&e^{-x} \mathbb{J}_{02}= \frac{3}{8 a^2} \left[\beta+\frac{16 (d-3) \alpha}{d-2}\right]-\frac{a^2 \Lambda }{\kappa}+\frac{1}{2 \kappa}\, ,\\
    \nonumber \mathbb{J}_{11}&=&  \frac{1}{8a^2} \left[\beta+\frac{4 (17 d-20) (d-3) \alpha}{(d-2) (d-1)}\right]+\frac{ a^2 \Lambda }{\kappa}+\frac{1}{2\kappa} \, ,\\
    \nonumber \mathbb{J}_{22}&=&  \frac{e^{2 x} \left\{(-8 a^4  \Lambda +12 a^2 )(d-2) (d-1)+\kappa(d-3)  [7 \beta d+4 (31 d-32) \alpha]+14 \beta \kappa\right\}}{16 a^2 (d-2) (d-1) \kappa}\, ,\\
 \mathbb{J}_{33}&=& \frac{1}{8 a^2}\left[\beta-\frac{8 \alpha}{d-1}\right]+\frac{a^2 \Lambda }{\kappa}+\frac{1}{2 \kappa}\, .
\end{eqnarray}

In order to illustrate the G\"{o}del solution, let us choose as the matter content, a perfect fluid, whose energy-momentum tensor is defined as
\begin{equation}
\label{17}
T_{\mu \nu} = (\rho + p)u_{\mu} u_{\nu} - p g_{\mu \nu}, 
\end{equation}
where $u_\mu$ is the four-velocity given by
\begin{equation}
\label{18}
u^{\mu} = (1,0,0,0)  ,\quad u_{\mu} = (a, 0 , a e^{x}, 0). 
\end{equation}
In this case, making use of the results \eqref{16} and source \eqref{17} the field equations \eqref{2} are cast as
\begin{eqnarray}
\label{20}
    \nonumber\frac{3}{8 a^2} \left[\beta+\frac{16 (d-3) \alpha}{d-2}\right]-\frac{a^2 \Lambda }{\kappa}+\frac{1}{2 \kappa} &=& a^2\rho \, ,\\
    \nonumber  \frac{1}{8a^2} \left[\beta+\frac{4\alpha  (17 d-20) (d-3) }{(d-2) (d-1)}\right]+\frac{ a^2 \Lambda }{\kappa}+\frac{1}{2\kappa}  &=&a^2 p \, ,\\
    \nonumber  \frac{   (12 a^2 -8 a^4  \Lambda )(d-2) (d-1)+\kappa (d-3)  [7 \beta d+4 (31 d-32) \alpha]+14 \beta \kappa }{16 a^2 (d-2) (d-1) \kappa} &=& \frac{1}{2} a^2   (p+2 \rho )\, ,\\
     \frac{1}{8 a^2}\left[\beta-\frac{8 \alpha}{d-1}\right]+\frac{a^2 \Lambda }{\kappa}+\frac{1}{2 \kappa} &=& a^2 p\, .
\end{eqnarray}

Although the system of equations \eqref{20} cannot be solved directly — meaning that no consistent solutions for $\rho$ and $\Lambda$ are found which would allow a direct comparison with the general relativity results for the G\"{o}del metric \cite{Godel:1949ga,Reboucas:1982hn} — insights of the general structure can be obtained by restricting to some particular cases.
Specifically, by setting $\alpha = 0$ (i.e. removing the Weyl tensor contributions) and $d = 4$, we obtain the following solution:
\begin{eqnarray}
\label{21}
    \rho&=& \frac{2 a^2+\beta  \kappa-2 a^4 \kappa p}{2 a^4 \kappa} \, ,\\
    \label{22}
     \Lambda &=& -\frac{-8 a^4 \kappa p+4 a^2+\beta  \kappa}{8 a^4 } \, .
\end{eqnarray}
The G\"{o}del original solution is recovered by taking $\beta = 0$ and considering a dust matter ($p = 0$).

It seems that the conformal sector (manifest in terms of the Weyl tensor) prevents the existence of G\"odel solutions.
Nonetheless, although we have no contribution from the conformal sector, an important feature of these solutions must be highlighted: it allows the possibility of violating the Weak Energy Condition (WEC) of general relativity when the inequality $2a^2 + \beta \kappa < 2a^4 \kappa p$ holds. This type of violation does not occur in standard general relativity, making it a distinctive feature of the modified framework considered here.

On the other hand, for the general case $\alpha \neq 0$, we attempted to determine solutions for a general dimension $d$ and for specific values $d = 3,4,5$, but no consistent solution was found.
Furthermore, going to a different class of model, keeping $\alpha \neq 0$ and setting $\beta = 0$, further attempts were made for unspecified $d$, as well as for $d = 3,4,5$, again without success — even in limiting cases such as $p = 0$ (pressureless matter) or $\rho = 0$ (vacuum).

These results strongly suggest that introducing the quadratic correction in the Weyl tensor into the gravitational action prevents the existence of G\"{o}del solution  as an exact solution of the modified gravity model.
In other words, this correction seems to rule out solutions like G\"{o}del's that allow for the existence of closed timelike curves (CTCs). Let us now consider the G\"{o}del-type solution.

\section{G\"{o}del-type universe}
\label{sec4}

In order to gain further insights about the conformal sector, we shall consider a wider class of metrics known as G\"{o}del-type solutions \cite{Reboucas:1982hn,Teixeira:1985wd} described by
\begin{equation}
\label{23}
    \mathrm{d}s^2 = [dt+H(r)d\phi]^2-D(r)^2 d \phi ^2 -dr^2 - dz^2,
\end{equation}
where the functions $H(r)$ and $D(r)$ are arbitrary functions of the radial coordinate.
These metric functions satisfy the conditions 
\begin{align} \label{23a}
\frac{H'\left(r\right)}{D\left(r\right)}	&=2\omega, \\
\frac{D''\left(r\right)}{D\left(r\right)}	&=m^{2},
\end{align}
so that the G\"odel-type metrics attain spacetime homogeneity \cite{Reboucas:1982hn,Teixeira:1985wd}.
The set of parameters $(\omega,m^2)$ completely characterizes all spacetime homogeneous G\"odel-type metrics.
In one hand, the parameter $\omega$ is the vorticity, while $m^2$ can assume any real value, $-\infty \leq m^2 \leq \infty$.
Among the allowed classes for this type of metric, we shall focus on the hyperbolic class \cite{Reboucas:1982hn,Teixeira:1985wd}, in which $m^2 >0$ and
\begin{eqnarray}
\label{24}
    H(r) &=& \frac{4 \omega}{m^2}\, \sinh ^2 \left( \frac{mr}{2}\right)\, ,\\
\label{25}    
    D(r) &=& \frac{1}{m}\, \sinh(mr)\, .
\end{eqnarray}
In regard to the presence of closed time-like curves (CTCs), we observe that for the hyperbolic class of such metrics, there exists a non-causal region $r>r_c$, in which the critical radius $r_c$ is given by
\begin{equation}
\sinh^2 \left(\frac{mr_c}{2}\right) = \left(\frac{4\omega^2 }{m^2}-1\right)^{-1}
\end{equation}
as long as $0< m^2< 4\omega^2$ \cite{Reboucas:1982hn,Teixeira:1985wd}.
Moreover, when $m^2 \geq 4\omega^2$ there is no acausal region, and CTCs are not present; in the special point $m^2 = 4\omega^2$, the critical radius $r_c \to \infty$.

The scalar invariants for the metric \eqref{23} are
\begin{eqnarray}
\label{}
R &=& g^{\mu \nu}R_{\mu \nu} = g_{\mu \nu}R^{\mu \nu} = 2 \left(m^2-\omega^2\right)\, ,\\
 R_{\mu \nu}R^{\mu \nu} &=& 2 \left(m^4-4 m^2 \omega^2+6 \omega^4\right)\, .
\end{eqnarray}
Once again, once $R$ is simply a constant, the derivative terms $\nabla_{\mu} \nabla_{\nu} R$ and $g_{\mu \nu} \nabla^{\rho} \nabla_{\rho} R$ on the field equation \eqref{2} do not contribute.

Furthermore, let us now focus on the Weyl tensor contributions.
For $d = 4$, the non-vanishing components of the tensor $B_{\mu \nu}$ \eqref{4} are given by
\begin{eqnarray}
\label{}
  \nonumber B_{00} &=& -3 \omega^2 \left(m^2-4 \omega^2\right)\, ,\\
  \nonumber B_{02} &=& -\frac{12 \omega^3 \left(m^2-4 \omega^2\right) \sinh ^2\left(\frac{m r}{2}\right)}{m^2}\, ,\\
  \nonumber B_{11} &=& -\frac{1}{12} \left(m^2-4 \omega^2\right) \text{csch}^2(m r) \left[\left(4 m^2+15 \omega^2\right) \cosh (2 m r)+4 m^2-8 \omega^2 \cosh (m r)-7 \omega^2\right]\, ,\\
   B_{22} &=& -\frac{3 \omega^2 \left(m^2-4 \omega^2\right) \sinh ^2\left(\frac{m r}{2}\right) \left[\left(m^2+8 \omega^2\right) \cosh (m r)+m^2-8 \omega^2\right]}{m^4}\, ,
\end{eqnarray}
while the non-vanishing components of the tensor $Z_{\mu \nu}$ \eqref{5} are
\begin{eqnarray}
\label{}
    \nonumber Z_{00} &=& \frac{1}{3} \left(m^4-6 m^2 \omega^2+8 \omega^4\right)\, , \\
    \nonumber Z_{12} &=& \frac{4 \omega \left(m^2-4 \omega^2\right) \left(m^2-2 \omega^2\right) \sinh ^2\left(\frac{m r}{2}\right)}{3 m^2}\, , \\
    \nonumber Z_{11} &=& \frac{1}{3} \left(m^4-7 m^2 \omega^2+12 \omega^4\right)\, , \\
    \nonumber Z_{22} &=& \frac{2 \left(m^2-4 \omega^2\right) \sinh ^2\left(\frac{m r}{2}\right) \left(m^4-7 m^2 \omega^2+\left(m^4+m^2 \omega^2-8 \omega^4\right) \cosh (m r)+8 \omega^4\right)}{3 m^4}\, , \\
     Z_{33} &=& -\frac{1}{3} \left(m^2-4 \omega^2\right)^2\, . 
\end{eqnarray}
Since the contributions from the Weyl tensor do not vanish, we observe that the G\"odel-type universe are not conformally flat.

Therefore, with these results, we can compute the non-vanishing components of the tensor $\mathbb{J}_{\mu \nu}$ (which gather all geometric information from the field equation \eqref{2}), which are given by
\begin{eqnarray}
\label{eq30}
    \nonumber \mathbb{J}_{00}&=&\frac{m^{2}}{4\omega\sinh^{2}\left(\frac{mr}{2}\right)}\mathbb{J}_{02} \cr
    &=&  \frac{\kappa m^4 (2 \alpha-3 \beta)-6 m^2 \left[\kappa \omega^2 (8 \alpha-3 \beta)+1\right]+5 \kappa \omega^4 (32 \alpha-3 \beta)-6 \Lambda +18 \omega^2}{6 \kappa}\, ,\\
    \nonumber \mathbb{J}_{11}&=& -\frac{ \alpha   \left(m^2-4 \omega^2\right) \text{csch}^2(m r) \left[3 \left(m^2+6 \omega^2\right) \cosh (2 m r)+5 \left(m^2-2 \omega^2\right)-8 \omega^2 \cosh (m r)\right]}{6 } \\
    \nonumber&&+\frac{ 2\Lambda +2 \omega^2- \beta \kappa \left(m^2-3 \omega^2\right) \left(m^2-\omega^2\right)}{2\kappa}\, ,\\
    \nonumber \mathbb{J}_{22}&=& \frac{\sinh ^2(\frac{m r}{2})}{3\kappa m^4} \biggl \{ 2 \alpha \kappa (m^2-4\omega ^2 )   \left(m^4-16 m^2 \omega ^2+80 \omega ^4\right) \cr
    && +\left(m^2-4 \omega ^2\right) \cosh (m r) \left\{\kappa \left[m^4 (2 \alpha-3 \beta)-4 m^2 \omega ^2 (4 \alpha+3 \beta)+5 \omega ^4 (3 \beta-32\alpha)\right]+6 \left(\Lambda -3 \omega ^2\right)\right\} \\
    \nonumber&& -3 \beta \kappa (m^2-\omega ^2 )  \left(m^4-7 m^2 \omega ^2+20 \omega ^4\right) +6 m^2 \left(\Lambda +5 \omega ^2\right)+24 \omega ^2 \left(\Lambda -3 \omega ^2\right) \biggr \} \, ,\\
      \mathbb{J}_{33}&=& \frac{1}{6} \left[3 \beta \left(m^2-\omega^2\right)^2-2 \alpha \left(m^2-4 \omega^2\right)^2\right]+\frac{\Lambda +m^2-\omega^2}{\kappa}\, . 
\end{eqnarray}

Since the geometric part of the field equations is established for the G\"odel-type metric, we shall consider now different types of matter content to explore physically viable solutions of the field equations.

\subsubsection{Perfect fluid}
\label{4.1}

The energy-momentum tensor of a perfect fluid is defined in eq.~\eqref{17}, in which the four-velocity is defined as usual $u^{\mu} = \delta^{\mu}_0 =  (1, 0, 0, 0)$, while the corresponding covector is
\begin{equation}
\label{25}
    u_{\mu}= \left(1,0,\frac{4 \omega \sinh ^2\left(\frac{m r}{2}\right)}{m^2},0\right) \, .
\end{equation}
Thus, the non-vanishing components of the energy-momentum tensor are
\begin{eqnarray}
\label{}
    \nonumber T_{00} &=& \rho\, ,\\ 
    \nonumber T_{02} &=& \frac{4 \omega \rho  }{ m^2} \sinh ^2 \left(\frac{m r}{2}\right) \, ,\\
    \nonumber T_{11} &=&T_{33}= p \, ,\\
     T_{22} &=& \frac{16 \omega^2 \rho   \sinh ^4 \left(\frac{m r}{2}\right)  -m^2 p  \sinh ^2 \left(m r\right) }{m^4} \, .
\end{eqnarray}

Once the framework is prepared, the next step is to search for solutions to the field equations $\mathbb{J}_{\mu\nu} = T_{\mu\nu}$.
First, we attempted general solutions, considering $d=4$ and arbitrary free parameters $\alpha$, $\beta$, $\omega$ and $m$. However, we found that the system of equations does not admit consistent solutions in this general case.
Actually, the only mathematically consistent solution corresponds to the case $\alpha = 0$ and $m^2 = 2\omega^2$. This last condition leads to the standard G\"{o}del solution, and we recover the same results for $\rho$ and $\Lambda$ presented in the previous section, given by Eqs. \eqref{21} and \eqref{22}, respectively.
Hence, no contribution from the conformal sector is obtained.

\subsubsection{Massless scalar field}

Another source known to engender interesting features in the G\"odel-type universe is a scalar field \cite{Reboucas:1982hn,Teixeira:1985wd}.
The energy-momentum tensor that describes the massless scalar field is given by
\begin{equation}
\label{}
\mathcal{T}^{\mu \nu  }= \partial^{\mu}\phi\,\partial^{\nu}\phi-g^{\mu \nu}\,\Big(\frac{1}{2}\,\partial^{\alpha}\phi\,\partial_{\alpha}\phi+V(\phi)\Big) \, .
\end{equation}

For simplicity, we assume that the scalar field depends only on the variable $z$, that is, $\phi = \phi(z)$. The non-vanishing components of the energy-momentum tensor are given by:
\begin{eqnarray}
\label{}
\nonumber \mathcal{T}_{00}&=& - \mathcal{T}_{11}=  \frac{1}{2} \phi'(z)^2-V(\phi)\, ,\\
\nonumber \mathcal{T}_{02}&=& -\frac{2 \omega \sinh ^2\left(\frac{m r}{2}\right) }{m^2}\left[2 V(\phi)-\phi'(z)^2\right]\, ,\\
\nonumber \mathcal{T}_{22}&=& \frac{ \left[m^2 \sinh ^2(m r)-16 \omega^2 \sinh ^4\left(\frac{m r}{2}\right)\right]}{2 m^4} \left[2 V(\phi)-\phi'(z)^2\right] \, ,\\
 \mathcal{T}_{33}&=& \frac{1}{2} \phi'(z)^2+V(\phi)\, .
\end{eqnarray}

From the first relation we found that $\mathcal{T}_{00}=-\mathcal{T}_{11} $ which, when incorporated with the field equation, we can conclude that
\begin{equation}
\label{}
    \mathbb{J}_{00}=-\mathbb{J}_{11} \, .
\end{equation}
However, examining the structure of the components $\mathbb{J}_{00}$ and $\mathbb{J}_{11}$ in eq.~\eqref{eq30} we observe that they differ.
Importantly, a stringent condition emerges from this relation, namely $m^2 = 4\omega^2$.
Actually, from our previous discussion, it is worth to highlight that this condition yields to an infinite critical radius, thereby preventing causality violation and resulting in a completely causal solution.

Once we have obtained some features about the G\"odel-type metric, namely $m^2 = 4\omega^2$, we can examine the behavior of the scalar field by solving the gravitational field equations.
Hence, substituting the condition $m^2 = 4\omega^2$ into all components of $\mathbb{J}_{\mu \nu}$ eq.~\eqref{eq30}, we arrive at
\begin{eqnarray}
\label{eq36}
   \mathbb{J}_{00}&=&\frac{\omega}{\sinh^{2}\left(r\omega\right)}\mathbb{J}_{02}=-\mathbb{J}_{11}=-\frac{\omega^{2}}{\sinh^{2}\left(r\omega\right)}\mathbb{J}_{22}=\frac{3\beta\kappa\omega^{4}-2\Lambda-2\omega^{2}}{2\kappa},\cr
 \mathbb{J}_{33}&=& \frac{9 \beta \kappa \omega^4+2 \Lambda +6 \omega^2}{2 \kappa}\, ,
\end{eqnarray}
An important remark is that the condition $m^2 = 4\omega^2$ removes every contributions related with the Weyl tensor, i.e. the components of $\mathbb{J}_{\mu \nu}$ do not depend on the parameter $\alpha$.

Finally, solving the field equations $\mathbb{J}_{\mu \nu}=T_{\mu \nu}$, under the condition $m^2 = 4\omega^2$, yields
\begin{equation}
\label{eq37}
    \phi '(z)^2=\frac{2 \omega ^2 +6 \beta \kappa \omega ^4}{\kappa} ={\rm const} \, ,
\end{equation}
and
\begin{equation}
\label{eq38}
    \Lambda = \frac{1}{2} \left[-3 \beta \kappa \omega^4+2 \kappa V(\phi)-4 \omega^2\right] \, .
\end{equation}

Solving the differential equation \eqref{eq37} for $\phi$  as a function of $z$, we obtain
\begin{equation}
\label{}
    \phi(z) = \pm \sqrt{\frac{2 \omega ^2 +6 \beta \kappa \omega ^4}{\kappa}} z \, 
\end{equation}
while the potential $V(\phi)$ in \eqref{eq38} takes the form
\begin{equation}
\label{}
    V(\phi)=\frac{3 \beta \kappa \omega^4+2 \Lambda +4 \omega^2}{2 \kappa}\, .
\end{equation}
It is worth noting that, for this class of solution, the scalar potential is constant.

In summary, about the G\"odel-type solution within quadratic curvature gravity \eqref{1}, it is important to note that the system of field equations naturally imposes the condition $m^2 = 4\omega^2$ as a requirement for the existence of an analytical solution.
Unfortunately, this condition eliminates all terms proportional to the parameter $\alpha$, thereby removing any contributions arising from the correction term involving the Weyl tensor.
As a result, although the solution does not includes Weyl tensor corrections, the G\"{o}del-type metric remains a valid solution of the gravitational model when the matter content is a scalar field. At last, this solution corresponds to a causal G\"{o}del universe, meaning that closed timelike curves (CTCs) are not allowed.

\subsection{General homogeneous spacetime}

Since the quadratic curvature gravity \eqref{1} does not possess, in general, the G\"{o}del-type universe as a physical solution, unless some additional conditions are assumed, which unfortunately implies that all Weyl tensor corrections are eliminated (being by the necessity of take $\alpha = 0$ or  $m^2 = 4\omega^2$), we shall now consider the general homogeneous metric \eqref{23} subject to the conditions \eqref{23a} without specifying the functional forms of functions $H(r)$ and $D(r)$.

Moreover, in order to solve the field equations in terms of the parameters $(\omega^2,m^2)$ to make contact with the G\"odel-type solutions, we use the conditions \eqref{23a} to obtain $H''=2\omega m^{2}D'$ and $\frac{H'''}{D}= 2\omega m^{2}$, and also $D'''=m^{2} D'$ and $\frac{D''''}{D} =m^{4}$.
Hence, considering the energy-momentum tensor of a perfect fluid eq.~\eqref{17} as source, the field equations read
\begin{align}
\frac{80\alpha\omega^{4}}{3}-\frac{5\beta\omega^{4}}{2}-\frac{\Lambda}{\kappa}-\frac{m^{2}}{\kappa}+\frac{3\omega^{2}}{\kappa}+\frac{\alpha m^{4}}{3}-\frac{\beta m^{4}}{2}-8\alpha m^{2}\omega^{2}+3\beta m^{2}\omega^{2}=\,&\rho, \\
\frac{\Lambda}{\kappa}+\frac{\omega^{2}}{\kappa}+\frac{1}{3}\alpha m^{4}-\frac{1}{2}\beta m^{4}-\frac{16}{3}\alpha m^{2}\omega^{2}+2\beta m^{2}\omega^{2}+16\alpha\omega^{4}-\frac{3}{2}\beta\omega^{4}=\,&p,\\
\frac{\Lambda}{\kappa}-\frac{\omega^{2}}{\kappa}-\frac{\alpha m^{4}}{3}+\frac{\beta m^{4}}{2}+\frac{8}{3}\alpha m^{2}\omega^{2}-\beta m^{2}\omega^{2}-\frac{16\alpha\omega^{4}}{3}+\frac{m^{2}}{\kappa}+\frac{\beta\omega^{4}}{2}=\,&p,
\end{align}
The general solutions for this set of equations, in terms of the parameters $(\omega^2,m^2)$ as well as $(\Lambda,\alpha, \beta)$, can be cast as the following
\begin{align}
\frac{2}{\kappa}\left(\omega^{2}-m^{2}\right)-\frac{4\Lambda}{\kappa}=\,&\rho-3p, \\
\frac{1}{\kappa}\left(4\omega^{2}-m^{2}\right)+\frac{2\alpha}{3} \left(4\omega^{2}-m^{2}\right)\left(16\omega^{2}-m^{2}\right)+\beta\left(4\omega^{2}-m^{2}\right)\left(m^{2}-\omega^{2}\right)=\,&\rho+p, \label{sol1}\\
-\frac{2\Lambda}{\kappa}+\frac{1}{\kappa}\left(2\omega^{2}-m^{2}\right)+\frac{8\alpha}{3} \omega^{2}\left(4\omega^{2}-m^{2}\right)+\beta\omega^{2}\left(m^{2}-\omega^{2}\right)=\,&\rho-p \label{sol2}
\end{align}
These expressions correspond to the field equations solutions of a general homogeneous spacetime for general functions $H(r)$ and $D(r)$ under the conditions \eqref{23a}.
These results are surprising, because when the usual G\"odel-type solutions was considered in sec.~\ref{4.1}, no physically viable solutions were obtained for $\alpha\neq 0$.

We observe from the equations \eqref{sol1} and \eqref{sol2} that in order to have nonvanishing Weyl term contributions  $\alpha\neq 0$ it is necessary that $m^2 \neq 4\omega^2$.
Hence, within the G\"odel-type universe, we still can discuss the presence of CTCs here depending whether $m^2 < 4\omega^2$ or $m^2>  4\omega^2$.
Moreover,  from the relation \eqref{sol1}, we see that if we remove the Weyl sector contributions with $m^2=4\omega^2$ we necessarily obtain a vacuum equation of state $\rho=- p$.

\section{Axially Symmetric Spacetime Metric}
\label{sec5}

Although we have seen interesting features of the G\"odel-type solutions within the quadratic curvature gravity \eqref{1}, every aspect of these solutions were independent of the Weyl tensor contributions, it seems that the conformal sector does not influences aspects about causality violation. 
Considering this fact, we have discussed the solutions of a general homogeneous spacetime  under the conditions \eqref{23a} and showed that in this universe, a new class of solutions with $\alpha\neq 0$ was obtained, and they contain very interesting aspects regarding the causality violation in the model \eqref{1}.
 
Moreover, to complement the analysis of the causal structure of the quadratic curvature gravity, we shall consider a different type of metric known to allows for the existence of CTCs and thus permits the violation of causality \cite{Haza, Ahmed, Ahmed1}.
It is a particularly interesting axially symmetric metric described by the line element
\begin{equation}
    \label{har}
ds^2 = \frac{dr^2}{\epsilon ^2 \, r^2} + r^2 dz^2 +\left (-2r^2 dt + \frac{\delta z }{r^2}dr-r^2 t\, d\phi \right)d\phi \, ,
\end{equation}
where $\epsilon$ and $\delta$ are nonzero constants, with $\delta > 0$.
The range of the variables are: $0 \leq r < \infty$, $-\infty < z < \infty$, $-\infty < t < \infty$, while $\phi$ is a periodic coordinate $\phi \sim \phi + \phi_{0}$, with $\phi_{0} > 0$.
It should be noted from Eq.~\eqref{har} that the spacetime has a coordinate singularity at $r = 0$ \cite{Haza, Ahmed, Ahmed1}.

It is interesting to highlight some aspects of the metric \eqref{har}: the condition under which closed curves arise comes precisely by imposing that $t = t_{0}$, $r = r_{0}$, and $z = z_{0}$, reducing the cylindrical spacetime to a line element in the form of a circle, with the form
\begin{equation}
    \label{}
    ds^2 = -r^2 t\, d\phi^2 \, .
\end{equation}
Therefore, there are three possible cases: if $t = t_0 > 0$, the trajectories are actually CTCs; for $t = t_0 = 0$, it is a closed null curve since it is a lightlike curve; and for $t = t_0 < 0$, it generates a closed spacelike curve.

Furthermore, it is possible to reduce the metric \eqref{har} to the case where $r = r_0$ and $z = z_0$, which leads to the line element describing Misner spacetime \cite{Misner:1965zz,Hawking:1973uf}
\begin{equation}
\label{}
ds^2 = -2r^2 dtd\phi-r^2 t\, d\phi^2 \, .
\end{equation}

The respective nonzero components of the Ricci tensor related with the metric \eqref{har} are
\begin{eqnarray}
    \label{}
    \nonumber R_{02}   = - R_{33}&=& 3 \epsilon ^2 r^2 \, ,\\
    \nonumber R_{11} &=& -\frac{3}{r^2} \, ,\\
    \nonumber R_{12}   &=& -\frac{3 \epsilon ^2 \delta z}{2 r^2} \, ,\\
      R_{22} &=& \frac{\epsilon ^2 \left(\delta ^2+24 r^6 t\right)}{8 r^4} \, .
\end{eqnarray}
Moreover, the scalar invariants for the metric \eqref{har} are
\begin{eqnarray}
    \label{eq45}
    \nonumber R=g_{\mu \nu}R^{\mu \nu}&=&-12 \epsilon ^2\, ,\\
    \nonumber R_{\mu \nu}R^{\mu \nu}&=& 36 \epsilon ^4 \, ,\\
     R_{\rho \sigma \mu \nu}R^{\rho \sigma \mu \nu}&=&24 \epsilon ^4 \, .
\end{eqnarray}
We observe that the Ricci scalar $R$ is just a constant, so that the derivative terms $\nabla_{\mu} \nabla_{\nu} R$ and $g_{\mu \nu} \nabla^{\rho} \nabla_{\rho} R$ do not contribute to the field equation \eqref{2}.
Moreover, the Kretschmann scalar in \eqref{eq45} shows that the apparent singularity at $r = 0$ is merely a coordinate singularity, that is, it is completely removable.

Let us now focus on the contributions from the Weyl tensor.
Moreover, we examine these contributions at arbitrary dimension $d$ in order to highlight some relevant aspects.
Now we will calculate the nonzero terms of the tensor $B_{\mu \nu}$ \eqref{4}
\begin{eqnarray}
\label{}
    \nonumber B_{02}&=& -\frac{4 \epsilon^4 \left(d^2-9 d+20\right) r^2}{d^2-3 d+2}\, , \\
    \nonumber B_{21}&=& -\frac{\epsilon ^2 \left(d^2-9 d+20\right)}{(d-2) (d-1) r}\, , \\
    \nonumber B_{22}&=& -\frac{\epsilon ^4 \left[\delta^2 \left(29 d^2-99 d+22\right)+12 \delta \left(d^2-7 d+14\right) r^3 z+64 \left(d^2-9 d+20\right) r^6 t\right]}{16 (d-2) (d-1) r^4}\, . 
\end{eqnarray}
It is important to note that for an arbitrary dimension $d$, the tensor $B_{\mu \nu}$ is asymmetric, because $B_{12} = 0$ and $B_{20} = 0$.
Actually, assuming the case $d = 4$, the tensor $B_{\mu \nu}$ becomes symmetric and the only nonzero component is
\begin{equation}
\label{}
    B_{22}= -\frac{\epsilon ^4 \delta \left(15 \delta +4 r^3 z\right)}{16 r^4}\,.
\end{equation}
Moreover, the tensor $Z_{\mu \nu}$ \eqref{5} is symmetric for any dimension, and the nonzero components are
\begin{eqnarray}
\label{}
    \nonumber Z_{02} &=& \frac{9 \epsilon^4 \left(d^2-9 d+20\right) r^2}{(d-2) (d-1)}\, , \\
    \nonumber Z_{11} &=& -\frac{9 \epsilon^2 \left(d^2-9 d+20\right)}{\left(d^2-3 d+2\right) r^2}\, , \\
    \nonumber Z_{12} &=& -\frac{9 \epsilon^4 \delta  \left(d^2-9 d+20\right) z}{2 (d-2) (d-1) r^2}\, , \\
    \nonumber Z_{22} &=& \frac{\epsilon^4 (d-4) \left[\delta ^2 (d+1)+36 (d-5) r^6 t\right]}{4 (d-2) (d-1) r^4}\, , \\
    \nonumber Z_{33} &=& -\frac{9 \epsilon^4 \left(d^2-9 d+20\right) r^2}{(d-2) (d-1)}\, .
\end{eqnarray}
Nonetheless, it is identically zero when $d = 4$, which is the case we are considering.

On the other hand, we can gather all the above geometric contributions of the field equation into the tensor $\mathbb{J}_{\mu \nu}$ yielding
\begin{eqnarray}
\label{}
    \nonumber \mathbb{J}_{02} &=& - \mathbb{J}_{33}= \frac{r^2 \left(\Lambda -3 \epsilon ^2\right)}{\kappa}\, , \\
    \nonumber \mathbb{J}_{11} &=& \frac{3 \epsilon ^2-\Lambda }{\epsilon ^2 \kappa r^2}\, , \\
    \nonumber \mathbb{J}_{12} &=& \frac{\delta z \left(3 \epsilon ^2-\Lambda \right)}{2 \kappa r^2}\, , \\
 \mathbb{J}_{22} &=& \frac{\epsilon ^2 \left(\delta^2-24 r^6 t\right)+8 \Lambda  r^6 t-\epsilon ^4 \delta \kappa \left(6 \delta \beta +15 \delta \alpha +4 \alpha  r^3 z\right)}{8 \kappa r^4}\, .
\end{eqnarray}

Moreover, we choose pure radiation as the matter content, as adopted in Refs.~\cite{Haza, Ahmed, Ahmed1}, whose energy–momentum tensor is given by
\begin{equation}
    \label{}
    T_{\mu \nu}=\rho \zeta _{\mu} \zeta _{\nu} \,,
\end{equation}
where the Killing vectors are
\begin{equation}
    \label{}
    \zeta ^{\mu} = (1,0,0,0) ,\quad \zeta _{\mu} = (0,0,-r^2,0). 
\end{equation}
The nonzero component of the energy–momentum tensor of the pure radiation field  is
\begin{equation}
    \label{}
    T_{22}=\rho  r^4 \, .
\end{equation}

Finally, solving the field equations for this type of matter, we obtain as a solution
\begin{eqnarray}
    \label{eq51}
    \rho &=& \frac{\epsilon^{2}\delta^{2}-3\epsilon^{4}\delta^{2}\kappa\left(2\beta+5\alpha\right)-4\epsilon^{4}\delta\alpha\kappa r^{3}z}{8 \kappa r^8}\, , \\
    \label{}
    \Lambda &=& 3 \epsilon ^2 \, .
\end{eqnarray}
For the first time, it is observed that the corrections due to the Weyl tensor contribute to the energy density.
Moreover, by taking $\kappa = 1$ and the parameters $\alpha$ and $\beta$ equal to zero, we recover the standard result of general relativity, i.e., $\rho = \frac{\epsilon ^2 \delta ^2}{8r^8}$ \cite{Haza, Ahmed, Ahmed1}.
Another important observation is that the cosmological constant has the opposite sign to the result obtained in general relativity. Thus, in this case, we have a change from an AdS universe ($\Lambda < 0$) to a dS universe ($\Lambda > 0$).

We can now establish important comments regarding the energy density: in general relativity, the result as seen above is that $\rho$ is always positive and decreases with increasing $r$, vanishing as $r \to \infty$.
On the other hand, in order to ensure that the Weak Energy Condition (WEC) from general relativity is not violated, we have the following conditions: $  \delta^{2} > 3\epsilon^{2}\delta^{2}\kappa\left(2\beta+5\alpha\right)+4\epsilon^{2}\delta\alpha\kappa r^{3}z$ for $z > 0$, and $ \delta^{2} + 4 \epsilon^{2} \delta \alpha k r^{3} z > 3\epsilon^{2}\delta^{2}\kappa\left(2\beta+5\alpha\right) $ for $z < 0$.
Although the density $\rho$ is a constant in time, its distribution changes spatially.

Another important aspect to consider is the term $4 \epsilon^4 \delta \alpha \kappa r^3 z$, which defines a critical radius for the violation of the WEC.
Although the solution is valid for all values of $r$, it is possible to identify which trajectories correspond to regions in which the WEC is violated.
To ensure that $\rho > 0$, assuming $\kappa = 1$, $\alpha > 0$, and $\beta > 0$, we obtain that
\begin{equation}
\label{}
    r < \biggl [\frac{\delta}{z}\biggl (\frac{1}{4\epsilon ^2 \alpha} -\frac{3 \beta}{2 \alpha} -\frac{15}{4}\biggr ) \biggr]^{\frac{1}{3}}\,,
\end{equation}
when $z > 0$, while for the case $z < 0$ we have
\begin{equation}
\label{}
    r > \biggl [\frac{\delta}{z}\biggl (\frac{1}{4\epsilon ^2 \alpha} -\frac{3 \beta}{2 \alpha} -\frac{15}{4}\biggr ) \biggr]^{\frac{1}{3}}\, .
\end{equation}
Observe that the situation is reversed when $\rho < 0$.

An interesting result of this analysis arises when
\begin{equation}
\label{}
r_* = \biggl [\frac{\delta}{z}\biggl (\frac{1}{4\epsilon ^2 \alpha} -\frac{3 \beta}{2 \alpha} -\frac{15}{4}\biggr ) \biggr]^{\frac{1}{3}} ,
\end{equation}
indicating a region with zero density, which can be interpreted as a critical radius separating regions with and without WEC violation.
Actually, the entire region satisfies the relation
\begin{equation}
\label{}
r^3\,z = \frac{\delta - 3 \epsilon^2 \delta (2 \beta + 5 \alpha)}{4 \alpha \epsilon^2} .
\end{equation}

Moreover, from the equation \eqref{eq51} we observe that the plane at $z = 0$ has a density
\begin{equation}
    \label{}
    \rho _{\rm flat }= \frac{\epsilon ^2 \delta ^2 -6 \epsilon ^4 \delta ^2 \beta -15 \epsilon ^4 \delta ^2 \alpha }{8  r^8}\, ,
\end{equation}
and for $\rho_{\rm flat} > 0$ it follows
\begin{equation}
\label{}
    1>3\epsilon ^2(2\beta+5\alpha).
\end{equation}

\begin{figure}[!htb]
    \centering
    \includegraphics[width=0.6\linewidth]{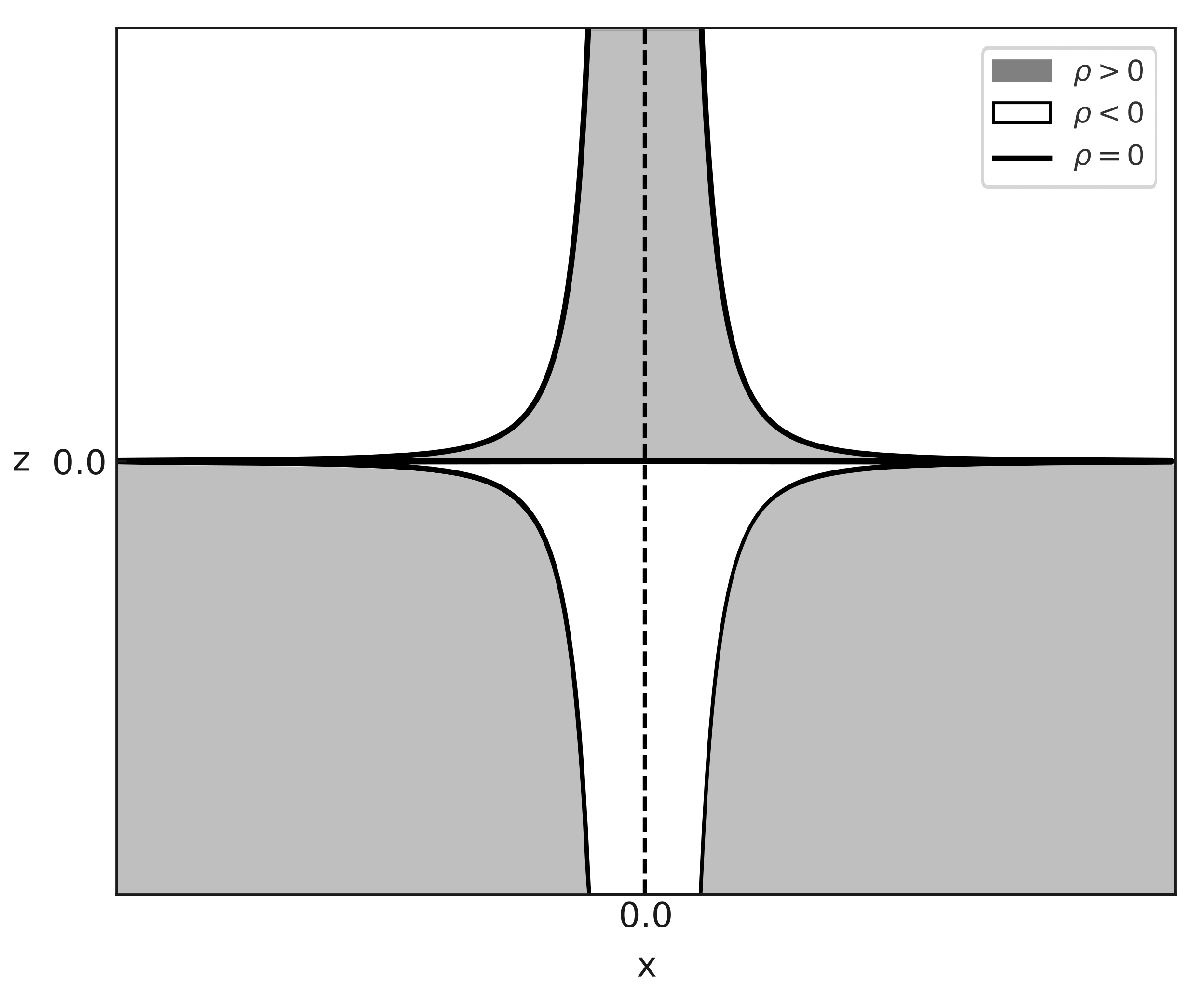}
    \caption{Energy density profile in the cylinder cross-section on the $Z \times X$ plane.}
    \label{fig1}
\end{figure}

Another relevant question is whether trajectories can access regions with $\rho < 0$, since locally this would not pose a problem in the reference frame of the object.
If there is a restriction, an object would need to cross the plane $z = 0$ to move from one semi-axis to the other, which would only be possible if the density in that plane satisfies $\rho_{\rm flat} > 0$.
A summary of these informations is shown in the figure \ref{fig1}.

One last point to examine is the rate of change of the energy density with the axial distance $z$. Hence, we obtain
\begin{equation}
\label{}
\frac{\partial \rho}{\partial z} = - \frac{\epsilon^4 \delta \alpha}{2 r^5}.
\end{equation}
This result indicates that the energy density decreases as $z$ increases, and that the rate of change is inversely proportional to the fifth power of the radius $r$.
In other words, the larger the value of $r$, the slower the density varies along the variable $z$.
This implies that the density is more sensitive to variations in $z$ near the central axis of the cylinder, while it becomes almost constant at larger radial distances.
It should be emphasized that this effect is exclusively due to the Weyl tensor (i.e. it depends on the parameter $\alpha$).

\section{Conclusions}
\label{sec6}

In this paper we have considered the possibility of causality violation solutions within the curvature squared gravity.
A point to emphasize is that the model under consideration (action and field equations) is written in terms of the Weyl tensor, which allowed to us to discuss aspects related with the conformal sector, mainly how it influences the presence of acausal solutions (i.e. closed timelike curves).

The first set of solutions we considered were the G\"odel and G\"odel-type universes.
In the G\"odel metric case we obtained explicitly the set of field equations, which surprisingly did not have a mathematical solution for the general case; but, insights could be established for some special cases.
By removing the Weyl tensor contributions and setting $d=4$ we could find a particular solution, which curiously allows the possibility of violating the WEC of general relativity, making it a distinctive feature.
However, it seems that the conformal sector (in terms of the Weyl tensor) prevents the existence of G\"odel solutions.

In order to try to understand the role played by the conformal sector in this type of models, we then considered the G\"odel-type metric.
First, when a perfect fluid was considered as source, no consistent solution is found for the field equations in the general case (with arbitrary parameters).
The only mathematically consistent solution yields $\alpha = 0$ and $m^2 = 2\omega^2$, which corresponds to the original G\"odel solution.
Moreover, when we choose a scalar field as source, it naturally generates  the condition $m^2 = 4\omega^2$, which prevents causality violation and results in a causal G\"odel universe.
Actually, this condition also removed every contribution related with the Weyl tensor to the solution.

Since the previous metric solutions were independent of the Weyl tensor contributions (i.e. conformal sector), we considered an axially symmetric metric known to allows the existence of CTCs.
In this case, by considering radiation as source, we finally managed to solve exactly the field equations for $\alpha \neq 0$.
Furthermore, we examined the possibility of WEC violation in terms of these solution, which yields to a series of constraints upon the parameters.
Interestingly, these constraints imposed bounds upon the radial variable, showing the existence of a critical radius $r_*$ separating regions with and without WEC violation.
One last remark is that rate of change of the energy density depends exclusively on the Weyl tensor, in which the density is more sensitive to variations in $z$ near the central axis.

It is very intriguing the behavior of the Weyl tensor (conformal sector) in preventing the G\"odel and G\"odel-type solutions.
Obviously, as shown by the solutions obtained for the axially symmetric spacetime, the conformal sector do not preclude the existence of acausal solutions.
However, it is not sufficiently clear the relation between the conformal sector and the causality violation, which demands a more detailed study about the impact of the conformal symmetry in the CTCs and the geometry of the spacetime metric.

\section*{Acknowledgments}

This work by A.F.S. is partially supported by National Council for Scientific and Technological
Development - CNPq project No. 312406/2023-1.
J.C.R. de  Souza thanks CAPES for financial support.
R.B. acknowledges partial support from Conselho
Nacional de Desenvolvimento Cient\'ifico e Tecnol\'ogico (CNPq Project No.~ 306769/2022-0).


\global\long\def\link#1#2{\href{http://eudml.org/#1}{#2}}
 \global\long\def\doi#1#2{\href{http://dx.doi.org/#1}{#2}}
 \global\long\def\arXiv#1#2{\href{http://arxiv.org/abs/#1}{arXiv:#1 [#2]}}
 \global\long\def\arXivOld#1{\href{http://arxiv.org/abs/#1}{arXiv:#1}}


\end{document}